\documentstyle[12pt]{article}

\def\be{\begin{equation}}
\def\ee{\end{equation}}
\def\ben{$$}
\def\een{$$}
\def\ba{\begin{array}{c}}
\def\ea{\end{array}}

\begin{document}

\titlepage
\vspace*{2cm}

 \begin{center}{\Large \bf
 Comment on ``Supersymmetry and Singular Potentials" by
 Das and Pernice [Nucl. Phys. B 561 (1999) 357]
  }\end{center}

\vspace{10mm}

 \begin{center}
Miloslav Znojil

 \vspace{3mm}

\'{U}stav jadern\'e fyziky AV \v{C}R, 250 68 \v{R}e\v{z}, Czech
Republic\footnote{e-mail: znojil@ujf.cas.cz}

\end{center}

\vspace{5mm}

\section*{Abstract}

Das and  Pernice [Nucl. Phys. B 561 (1999) 357 and arXiv:
hep-th/0207112] proposed that the Witten's supersymmetric quantum
mechanics may incorporate potentials with strong singularities
whenever one succeeds in their appropriate regularization. We
conjecture that one of the most natural recipes of this type results
from a detour to a non-Hermitian (usually called ${\cal PT}$
symmetric) intermediate Hamiltonian (with the real and discrete
spectrum) obtained by an infinitesimal complex shift of the
coordinate axis.

\vspace{5mm}

PACS   03.65.Fd; 03.65.Ca; 03.65.Ge; 11.30.Pb; 12.90.Jv

\newpage

\section{Introduction}

Our present remark is inspired by the very recent discussion of
Gangopadhyaya and Mallow \cite{GM} with Das and Pernice \cite{DPc}.
Although this discussion is purely technical by itself, it
re-attracts attention to the singularity paradox of Jevicki and
Rodriguez (JR, \cite{JR}) and to its role in Witten's supersymmetric
quantum mechanics (SUSYQM, \cite{Witten}). In particular, the
discussion re-opens the question of acceptability of the resolution
of the above JR paradox as offered by Das and Pernice in their older
and longer paper \cite{DP}. In such a setting we feel it useful to
re-tell the story and to put the whole problem under a new
perspective.

Let us start form the harmonic oscillator in one dimension which
plays the role of one of the most elementary illustrations and
realizations of the ideas of SUSYQM \cite{Khare}. It comes as a
definite surprise that after one restricts the same solvable
Hamiltonian to the mere half-line of coordinates, the
supersymmetrization immediately encounters severe technical
difficulties \cite{stat}. In their above-menioned remark,
Gangopadhyaya and Mallow \cite{GM} even tried to claim that the
supersymmetry (SUSY) of the latter (so called half-oscillator) model
becomes spontaneously broken. In their reaction, Das and Pernice
\cite{DPc} resolved the puzzle by detecting a subtle flaw in the
complicated construction of ref. \cite{GM}. They re-confirmed their
older conjecture and belief \cite{DP} that all the paradoxes of the
JR type may be resolved via a proper regularization of the
singularities in the potential in question.

In constructive manner one may recollect that the standard SUSYQM
considerations start from a factorizable ``first" Hamiltonian $H_F =
B \cdot A $, assigning to it its ``second" or ``supersymmetric"
partner $H_S = A \cdot B$. In the case of the so called unbroken
SUSY, the respective spectra are closely related, $E_{S,0}=E_{F,1}$,
$E_{S,1}=E_{F,2}$ etc. This means that the ``second" spectrum is
obtained as an upward shift of the ``first" one,
 \be
 E_{F,n}\ <\ E_{F,n+1}\ =\ E_{S,n}, \ \ \ \ n = 0, 1, \ldots \ .
 \label{iso}
 \ee
The ``first" ground state $E_{F,0}=0$ is exceptional and it does not
possess any SUSY partner.  The most transparent illustration of the
scheme (cf. the review \cite{Khare} for more details) is provided by
the two harmonic oscillators $H_F^{(HO)} = p^2+x^2-1$ and $H_S^{(HO)}
= p^2+x^2+1$ in the units $\hbar  = 2m = 1$ and with
 \be
 E_{F,0}^{(HO)}=0 < E_{F,1}^{(HO)} = E_{S,0}^{(HO)}= 4
 < E_{F,2}^{(HO)} = E_{S,1}^{(HO)}= 8 <
 \ldots \ .
 \label{HOiso}
 \ee
As we mentioned above, a nontrivial observation has been made by
Jevicki and Rodrigues \cite{JR} who emphasized that the mathematical
consistency of the theory seems to require the absence of
singularities in the factors $A$ and/or $B$. In their singularly
factorized example $H_F^{(sing.)} = p^2+x^2-3=B^{(sing.)} \cdot\
A^{(sing.)}$ one encounters a negative ground state energy in the
spectrum,
 \be
 E_{F,0}^{(sing.)} = -2, \
 E_{F,1}^{(sing.)} = 0, \
 E_{F,2}^{(sing.)} = 2, \
 E_{F,3}^{(sing.)} = 4, \
 \ldots \
 \label{HOen}
 \ee
and the emergence of a centrifugal-like singularity in its SUSY
partner, $ H_S^{(sing.)} = p^2+x^2-1+2/x^2 =A^{(sing.)} \cdot
B^{(sing.)}$. The main difficulty appears when we compare the
``second", $p-$wave spectrum
 \be
 E_{S,0}^{(sing.)} = 4,\
 E_{S,1}^{(sing.)} = 8,\
 E_{S,2}^{(sing.)} = 12,\
 \ldots \
\label{second}
 \ee
with its predecessor (\ref{HOen}).  Obviously, the characteristic
SUSY isospectrality (\ref{iso}) is lost.

The message is clear: One may either accept the regularity rule or
postulate a suitable regularization of $A^{(sing.)}$ and
$B^{(sing.)}$. Das and Pernice \cite{DP} paid attention to the latter
possibility.

\section{A broader context: Harmonic oscillator in more dimensions
and its textbook, ``hidden" regularization \label{jedna}}

Apparently, a simple-minded resolution of the Jevicki-Rodrigues
paradox is not difficult: One only has to replace the one-dimensional
interpretation of $H_F^{(sing.)} = p^2+x^2-3=B^{(sing.)}\cdot
A^{(sing.)}$ by its radial-equation version with $x \in (0,\infty)$
(and with the vanishing angular momentum $\ell$ of course). In this
way, both the SUSY-partner Hamiltonians become defined on the same
interval of coordinates and the original full-axis energies
(\ref{HOen}) become replaced by the usual $s-$wave spectrum
 \be
 E^{\ell=0}_{F,0} = 0, \
 E^{\ell=0}_{F,1} = 4, \
 E^{\ell=0}_{F,2} = 8, \
 \ldots \ .
\label{swave}
 \ee
The standard SUSY pattern is restored.  Unfortunately, once we move
beyond the present elementary example the problem of singularities
recurs \cite{Khare}. For this reason, Das and Pernice \cite{DP} have
proposed that within the general SYSYQM, all the singular SUSY
factors $A$ and $B$ should be treated via a suitable regularization.
This philosophy is generally accepted at present \cite{tata,zno}.

Before we proceed further, let us return once more to the above
innocent-looking $s-$wave solutions (\ref{swave}) where $\ell=0$ in
the radial Schr\"{o}dinger equation
 \be
\left [
-\frac{d^2}{dr^2} + r^2 +\frac{\ell(\ell+ 1)}{r^2}
\right ] \psi^{(\ell)}(r)
=E\, \psi^{(\ell)}(r), \ \ \ \ \ \ \ \ \ r \in (0,\infty).
\label{SE}
 \ee
The explicit parabolic-cylinder \cite{GM} wave functions
$\psi^{\ell=0}(r)$ of this equation are chosen as asymptoticaly
decreasing at any real value of the energy parameter $E>0$. In a
naive but fairly popular setting (a sketchy review of which may be
found elsewhere \cite{ZoSD}) one then requires the normalizability of
the wave function near the origin, i.e., its threshold behaviour
$\psi^{(\ell)}(r) \sim r^{-1/2+\delta}$ with a suitable $\delta > 0$.
To one's great surprize, this condition guarantees the discrete
character of the spectrum for the sufficiently large $\alpha = \ell
+1/2>1$ only.  Thus, for our regular $V(r) = r^2$ in particular, the
quantization follows from the normalizability only for $\ell = 1, 2,
\ldots $ in eq. (\ref{SE}). The $s-$wave (or, in general, any real
$\alpha=\ell+1/2 \in (0,1)$) is exceptional and its quantization
requires an {\em additional} boundary condition. This requirement is
based on the {\em deeply physical} reasoning \cite{LL} and
represents, in effect, just an {\em independent} postulate of an
appropriate regularization
 \be
\left \{
 \begin{array}{c}
\psi^{(\ell)}(0) = 0\\
\lim_{r \to 0} \psi^{(\ell)}(r)/\sqrt{r} = 0
 \ea
\right .
\ \ \ \ \
{\rm for}\ \ \  \
\left \{
 \begin{array}{c}
 \alpha \in [1/2,1) \\
\alpha \in (0,1/2)
 \ea
\right . \label{regu}
 \ee
in quantum mechanics. This point is highly instructive and its
importance is rarely emphasized in the textbooks where the
independence of the regularization (\ref{regu}) is usually denied
(for example, the Newton's \cite{Newton} proof of eq. (\ref{regu})
holds for $\ell=0$ only) or disguised (for example, the Fl\"{u}gge's
\cite{Fluegge} requirement of the boundedness of the kinetic energy
{\em is} in effect a new, independent postulate).

In such a context it is not too surprizing that the suppression
(\ref{regu}) of the subdominant components of $\psi^{(\ell)}(r)$ near
$r=0$ is sometimes being replaced by an alternative requirement.  The
half-oscillator regularization of refs. \cite{DP,GM,DPc} offers one
of its most characteristic examples. In a more formal mathematical
setting (see, e.g., the Reed's and Simon's monograph \cite{RS}), one
should of course use a more rigorous language and replace the word
``regularization" by the phrase ``selection of a suitable essentially
self-adjoint extension" of the Hamiltonian operator in question.

\section{Regularization recipe of Das and Pernice \label{two}}

The necessity of a restoration of SUSY in eq. (\ref{SE}) (let us just
take now $\ell(\ell+1)=0$ for simplicity) has led the authors of ref.
\cite{DP} to an artificial extension of the range of the coordinates
to the whole axis, $r \in (-\infty,\infty)$. This was compensated by
an introduction of a ``very high" barrier to the left, $V(x) = c^2
\gg 1$ for $r \in (-\infty,0)$.  Under this assumption they succeeded
in a reconstruction of SUSY partnership between the two
harmonic-oscillator-like potentials
 \be
\ba V_F(x) = (x^2-1)\,\theta(x) + c^2\theta(-x) - c\,\delta(x),\\
V_S(x) = (x^2+1)\,\theta(x) + c^2\theta(-x) + c\,\delta(x).
 \ea
\label{DPP}
 \ee
Both of them depend on the sign of $x$ (via Heavyside functions
$\theta$) and contain Dirac delta-functions multiplied by the
above-mentioned large constant $c \gg 1$.

The above-mentioned, asymptotically correct parabolic-cylinder wave
functions $\psi^{(\ell=0)}(x) $ still satisfy the corresponding
Schr\"{o}dinger equation whenever $x> 0$. The price to be paid for
the presence of the delta-function in eq. (\ref{DPP}) lies in a
violation of the boundary condition (\ref{regu}). In place of the
elementary $\psi^{(\ell=0)}(0) = 0$ we now have the two separate and
more complicated requirements
 \be
\left \{
 \begin{array}{c}
\mu_F(E,c)\,\psi^{(\ell=0)}_F(0)+\nu_F(E,c)\,
\psi'^{(\ell=0)}_{F}(0) = 0\\
\mu_S(E,c)\,\psi^{(\ell=0)}_S(0)
+\nu_S(E,c)\,\psi'^{(\ell=0)}_{S}(0) = 0
 \ea
\right . .\label{SUSYregu}
 \ee
They mix the values $\psi(0)$ and derivatives $\psi'(0)$ of the wave
function in the origin. The explicit form of the coefficients $\mu$
and $\nu$ has been given in refs.~\cite{GM,DPc} as well as \cite{DP}.
In the special case of the half-oscillator limit $c \to \infty$,
equation (\ref{SUSYregu}) degenerates to the much simpler rule
 \be
\left \{
 \begin{array}{c}
\psi'^{(\ell=0)}_{F}(0) = 0\\
\psi^{(\ell=0)}_S(0) = 0
 \ea
\right . .
\label{finSUSYregu}
 \ee
We may summarize that in our above notation and units, the $c \to
\infty$ recipe is based, simply, on a sophisticated replacement of
the full-line spectrum (\ref{HOen}) by its subset selected by the
Neumann boundary condition (\ref{finSUSYregu}) in the origin,
 \be
 E^{(DP)}_{F,0} = 0, \
 E^{(DP)}_{F,1} = 4, \
 E^{(DP)}_{F,2} = 8, \
 \ldots \ .
 \label{DPHOen}
 \ee
As long as the ``second" spectrum (\ref{second}) remains unchanged by
the Dirichletian second line in eq. (\ref{finSUSYregu}), the two SUSY
partner spectra obey, by construction, the unbroken SUSY and its
isospectrality rule (\ref{iso}) even in the limit $c\to \infty$. The
mathematics has got simplified -- all we need are just the adapted
boundary conditions (\ref{finSUSYregu}). Of course, within the
textbook quantum mechanics their intuitive physical acceptability is
less obvious due to their manifest disagreement with the much more
common suppression (\ref{regu}) of dominant terms.

\section{SUSY via analytic continuation
\label{three}}

Buslaev and Grecchi \cite{BG} were probably the first who understood
that a complex shift of coordinates may leave the spectrum (or at
least its part) in many radial Schr\"{o}dinger equations unchanged.
In their spirit we introduced the complexified radial oscillator
 \be
\left [ -\frac{d^2}{dx^2} + (x-i\,\varepsilon)^2 +\frac{\ell(\ell+
1)}{ (x-i\,\varepsilon)^2} \right ] \psi^{(\ell,\varepsilon)}[r(x)]
=E\, \psi^{(\ell,\varepsilon)}[r(x)]
 \label{SEBG}
 \ee
with $x \in (-\infty,\infty)$ and analyzed its spectrum under the
standard pair of the asymptotic boundary conditions
 \be
\lim_{X \to \pm \infty} \ \psi^{(\ell,\varepsilon)}[r(X)] \ = \ 0.
\label{SEBGbc}
 \ee
The work has been done in ref. \cite{PTHO} and its result may be
perceived, in SUSY context, as a regularization recipe. It is worth
noticing that the new recipe is different from the very specific and
closely SUSY-related technique of preceding section.

One of the most distinguished features of the use of the complex
shift $\varepsilon > 0$ in eqs. (\ref{SEBG}) and (\ref{SEBGbc}) is
that it produces the discrete energies
 \be
 E^{(\pm)}_{BG,N}(\alpha) = 4N+2 \mp 2\alpha
 \label{ptHOen}
 \ee
which are all real and numbered by the integers $N = 0, 1, \ldots$
{\em and} by the (superscripted) quasiparities $Q = \pm$. In this
way, the complexification reproduces the one-dimensional
harmonic-oscillator spectrum at $\ell =\alpha - 1/2=0$ and extends it
in a certain non-empty vicinity of $\alpha = 1/2$ in an entirely
smooth manner.

One may use the similar analytic continuation technique within the
SUSYQM context as well. As a universal regularization recipe, it has
been proposed in ref. \cite{zno}. Its consequences proved extremely
satisfactory in the domain of $\varepsilon > 0$ (we may refer to {\it
loc. cit.} for more details).

What happens when we perform the backward limiting transition
$\varepsilon \to 0$?  Firstly, the de-complexification of the
coordinate $r(x)$ splits the whole real line of $x$ in two (viz.,
positive and negative) half-axes of $r$ which become completely
separated \cite{ZoSD}. Fortunately, there occur only marginal
discontinuities in the spectrum itself.  Manifestly, this is
illustrated in Figure 1 where

\begin{itemize}

\item
at a fixed $N$, the ordering of the levels
 \ben
E_{F}^{(+)} \leq  E_{S}^{(+)}  \leq  E_{F}^{(-)} \leq E_{S}^{(-)}
\label{ordered}
 \een
is preserved for all the values of an auxiliary real $\gamma \in
I\!\!R $ which parametrizes both $\alpha_F = |\gamma|$ and $\alpha_S
= |\gamma+1|$;

\item
each of the above four energies is a piecewise linear function of
$\gamma$ which changes its slope at a single point;

\item
the domain of $E_{F}^{(+)}$ or $ E_{S}^{(+)}$ is a finite interval,
out of which the related wave function $\psi$ ceases to belong to the
Hilbert space in the limit $\varepsilon \to 0$.

\end{itemize}

\noindent The complete isospectrality of $H_F$ and $H_S$ occurs far
to the left and right in the picture,
 \ben
\left \{
\begin{array}{lll}
E_{F}^{(-)}(N) \ = \ E_{S}^{(-)}(N), &\gamma \in (-\infty,-2),& \\
E_{F}^{(-)}(N+1) \
 =
E_{S}^{(-)} (N),& \gamma \in (1, \infty), &
 N \geq 0. \ea \right .
 \een
In the domain of the upper line the ground-state energy is positive
and all the spectrum is $\alpha-$dependent and equidistant.  In
contrast, in the domain of the lower line, the SUSY is unbroken and
the spectrum is $\alpha-$independent.

In the non-central interior intervals, the SUSY-type isospectrality
is destroyed by the new levels which start to exist. The SUSY
interpretation of the partners is lost in the intervals $\gamma \in
(-2,-1)$ and $\gamma \in (0,1)$.  A manifest breakdown of SUSY is
encountered in this regime.

The most interesting pattern is obtained in the central interval of
$\gamma \in (-1,0)$.  The completely standard SUSY behaviour is
revealed there for all $\alpha_F = 1-\alpha_S$. The SUSY-related
isospectrality holds there, with the exceptional ground state
$E_F^{(+)}(0)$. At a fixed main quantum number $N$ we have
 \ben
E_{F}^{(+)} \ <\  E_{S}^{(+)} \ =\ E_{F}^{(-)} \ <\
   E_{S}^{(-)}
   \een
and return to eq. (\ref{HOiso}) at $\alpha = 1/2$. The SUSY is
unbroken even though the spectrum itself ceases to be equidistant at
$\alpha \neq 1/2$.

\section{Conclusions}

We may summarize that the SUSY regularization via an intermediate
complexification of ref. \cite{zno} represents an improved
implementation of the universal idea of ref. \cite{DP}. First of all,
it avoids the incompleteness  of the supersymmetrization performed in
the spirit of section \ref{two} which works, as a rule, just with a
subset of the whole one-dimensional spectrum. In contrast, the
complexification recipe (with its subsequent $\varepsilon \to 0$
return to the real axis) works with the {\em complete}
one-dimensional spectrum. Moreover, it leads to a consequent and {\em
smooth} $\ell \neq 0$ generalization of the one-dimensional SUSY
scheme. In the other words, the regularization of section \ref{three}
remains applicable within all the interval of $\ell\in (-1/2,1/2)$
where our ``spiked" harmonic oscillator potentials (with various
interesting applications \cite{spiked}) are strongly singular.


\section*{Acknowledgement}

Work supported by the grant Nr. A 1048004 of GA AS CR.

\section*{Figure captions}

Figure 1. Degeneracy of the spectrum and its $\gamma-$dependence
after the SUSY regularization of section \ref{three}.


\end{document}